\documentclass[epj]{webofc}
\usepackage[utf8]{inputenc}
\usepackage[varg]{txfonts}   
\usepackage{booktabs}
\usepackage{xcolor}
\definecolor{darkred}{rgb}{0.4,0.0,0.0}
\definecolor{darkgreen}{rgb}{0.0,0.4,0.0}
\definecolor{darkblue}{rgb}{0.0,0.0,0.4}
\usepackage[bookmarks,linktocpage,colorlinks,
    linkcolor = darkred,
    urlcolor  = darkblue,
    citecolor = darkgreen]{hyperref}
\usepackage{subfigure}
\usepackage{ulem}
\usepackage{bm}
\wocname{EPJ Web of Conferences}
\woctitle{Lattice2017}
\usepackage{amsmath}	
\begin{document}
%
\selectlanguage{english}
\title{%
Nucleon structure from 2+1 flavor lattice QCD near the physical point
}
\author{%
\firstname{Natsuki}  \lastname{Tsukamoto}\inst{1}\fnsep\thanks{Speaker, 
\email{tsukamoto@nucl.phys.tohoku.ac.jp}
}
\and
\firstname{Ken-Ichi} \lastname{Ishikawa}\inst{2}\fnsep\inst{3}
\and
\firstname{Yoshinobu} \lastname{Kuramashi}\inst{4}\fnsep\inst{5}\fnsep\inst{3}
\and
\firstname{Shoichi} \lastname{Sasaki}\inst{1}\fnsep\inst{3}
\and
\firstname{Takeshi} \lastname{Yamazaki}\inst{5}\fnsep\inst{4}\fnsep\inst{3}
\;\;for PACS Collaboration
}
\institute{%
Department of Physics, Tohoku University, Sendai, Miyagi 980-8578, Japan
\and
Department of Physics, Hiroshima University, Higashi-Hiroshima, Hiroshima 739-8526, Japan
\and
RIKEN Advanced Institute for Computational Science, Kobe, Hyogo 650-0047, Japan
\and
Center for Computational Sciences, University of Tsukuba, Tsukuba, Ibaraki 305-8577, Japan
\and
Faculty of Pure and Applied Sciences, University of Tsukuba, Tsukuba, Ibaraki 305-8571, Japan
}

\abstract{%
We present an update on our results of nucleon form factors measured on
a large-volume lattice $(8.1\mathrm{fm})^4$ at almost the physical
point in 2+1 flavor QCD. The configurations are generated with the
stout-smeared $\mathcal{O}(a)$ improved Wilson quark action and Iwasaki
gauge action at $\beta=1.82$, which corresponds to the lattice spacing
of 0.085 fm. The pion mass at the simulation point is about 145 MeV. We
determine the iso-vector electric radius and magnetic moment
from nucleon electric ($G_E$) and magnetic ($G_M$) form factors. 
We also report on preliminary results of the axial-vector ($F_A$), induced pseudo-scalar ($F_P$)
and pseudo-scalar ($G_P$) form factors in order to verify the axial Ward-Takahashi identity 
in terms of the nucleon matrix elements, which may be called as the generalized Goldberger-Treiman relation. 
}
\maketitle
\section{Introduction}
In general, four form factors appear in the nucleon matrix elements
of the weak current. Here, for example, we consider the matrix element
for neutron beta decay. In this case, the vector and axial-vector currents
are given by $V^{+}_{\alpha}(x)=\bar{u}(x)\gamma_{\alpha}d(x)$ and
$A^{+}_{\alpha}(x)=\bar{u}(x)\gamma_{\alpha}\gamma_5d(x)$,
and then the general form of the matrix element for neutron
beta decay is expressed by both the vector and axial-vector transitions:
\begin{align}
\langle p|V_{\alpha}^{+}(x)+A_{\alpha}^{+}(x)|n\rangle=\bar{u}_p \left(
{\cal O}_{\alpha}^V(q)+{\cal O}_{\alpha}^A(q)
\right) u_n e^{iq\cdot x},
\end{align}
where $q=P_n -P_p$ is the momentum transfer between the neutron ($n$)
and proton ($p$). The vector ($F_V$) and induced tensor ($F_T$) form
factors are introduced for the vector matrix element:
\begin{align}
{\cal O}_{\alpha}^{V}(q)=\gamma_{\alpha}F_V(q^2)+\sigma_{\alpha \beta}q_{\beta}F_T(q^2)
\end{align}
and also the axial-vector ($F_A$) and induced pseudo-scalar ($F_P$) form factors for the axial-vector matrix element:
\begin{align}
{\cal O}_{\alpha}^{A}(q)=\gamma_{\alpha}F_A(q^2)+iq_{\alpha}\gamma_5
F_P(q^2).
\end{align}

The vector part of weak processes is related to the nucleon's electromagnetic 
matrix element through an isospin rotation if heavy flavor contributions are 
ignored {\it under the exact isospin symmetry} ($m_u=m_d$). The iso-vector 
electric $G_E(q^2)$ and magnetic $G_M(q^2)$ Sachs form factors are given by appropriate linear 
combinations of $F_V(q^2)$ and $F_T(q^2)$ as
\begin{align}
G_E(q^2)=F_V(q^2) - \frac{q^2}{2M_N}F_T(q^2)\;\;{\rm and}\;\; G_M(q^2)=F_V(q^2) + 2M_N F_T(q^2),
\end{align}
where $M_N$ denotes the nucleon mass. 
Both form factors $G_{E/M}(q^2)$ can be expanded in powers of $q^2$ as
 %
 %
 \begin{align}
  G_{E(M)}(q^2) =G_{E(M)}(0)\left(1 - \frac{q^2}{6} \langle r^2 \rangle_{E(M)} + \frac{q^4}{120}\langle r^4\rangle_{E(M)} +\mathcal{O}(q^6)\right),
 \end{align}
where the electric (magnetic) root-mean-square (RMS) radius $r_{E(M)}=\sqrt{\langle r^2 \rangle_{E(M)}}$ 
can be evaluated from the slope of the form factors at zero momentum transfer. The value of $G_E(0)$ ($G_M(0)$)
corresponds to a difference between charges (magnetic moments) of the proton and neutron states.

The latest lattice calculations of the nucleon structure have been carried out 
with increasing accuracy~\cite{Bhattacharya:2013ehc,Green:2014xba,Capitani:2015sba,Alexandrou:2013joa,Alexandrou:2017ypw}.
It seems that there is still a gap between experimentally known values and
the lattice results, especially for the electric-magnetic nucleon radii and the magnetic moment.   
Our preliminary results computed with almost physical pion mass on very large volume 
shows agreement with experimental results~\cite{Kuramashi:2016lql}. 
In this paper, we present an update of our previous study~\cite{Kuramashi:2016lql}.

For the axial-vector part of weak processes, it is known that
$F_A(q^2)$ and $F_P(q^2)$ are not fully independent. 
It is because the axial Ward-Takahashi identity: $\partial_\alpha A^+_\alpha(x)=2\hat{m}P^{+}(x)$
leads to the generalized Goldberger-Treiman  (GT) relation among three form factors~\cite{Sasaki:2007gw}: 
\begin{align}
2M_NF_A(q^2)-q^2F_P(q^2)=2\hat{m}G_P(q^2),
\label{Eq:GTrelation}
\end{align}
where $\hat m=	m_u=m_d$ is a degenerate up and down quark mass and
the pseudo-scalar nucleon form factor $G_P(q^2)$ is defined in 
the pseudo-scalar nucleon matrix element
\begin{align}
\langle p|P^{+}(x)|n\rangle = \bar{u}_p \left( \gamma_5 G_P(q^2)
\right) u_n e^{iq\cdot x}
\label{Eq:PSFF}
\end{align}
with a local pseudo-scalar density, $P^{+}(x)\equiv\bar{u}(x)\gamma_{5}d(x)$.
Therefore, the $q^2$ dependences of three form factors, $F_A(q^2)$, $F_P(q^2)$ and $G_P(q^2)$
are constrained by Eq.~(\ref{Eq:GTrelation}) as a consequence of the axial Ward-Takahashi identity.
In this work, we also report on preliminary results of 
three form factors, $F_A(q^2)$, $F_P(q^2)$ and $G_P(q^2)$
in order to verify the axial Ward-Takahashi identity in terms of the nucleon
matrix elements.
 
 \section{Simulation details}
 We use the 2+1 flavor QCD gauge configurations generated with
 the stout-smeared ${\cal O}(a)$ improved Wilson-clover quark
 action and the Iwasaki gauge action on a $L^3\times T=96^3\times 96$ lattice
 at $\beta$ =1.82, which corresponds to 
 a lattice cutoff of $a^{-1} \approx 2.3\mathrm{GeV}$ 
 ($a\approx 0.0085$ fm)~\cite{Ishikawa:2015rho}.
 The improvement coefficient, $C_{\rm SW}=1.11$, is determined 
 nonperturbatively by the Schr\"odinger functional (SF) scheme. 
 The hopping parameters for the light sea quarks 
 $\{ \kappa_{ud}, \kappa_s\}=\{0.126117, 0.124812\}$ 
 are chosen to be near the physical point.
 We calculate iso-vector nucleon form factors in the vector, axial-vector
 and pseudo-scalar channels with a large spatial volume of $(8.1\;{\rm fm})^3$ 
 and a simulated pion mass reaching down to $m_\pi = 145$ MeV in 2+1 flavor QCD.
 A brief summary of our simulation parameters are tabulated in Table~\ref{Tab:Summary_Sim}.

Our preliminary results of the nucleon form factors 
with less number of measurements were published in Ref.~\cite{Kuramashi:2016lql}.
The total number of gauge configurations used in the present work is 200
which corresponds to 2000 trajectories. Each measurement is separated 
by 10 trajectories. We use the jackknife method with a bin size of 50 trajectories for estimating the statistical errors.
For technical details of how to extract the nucleon form factor from three-point functions, 
see Refs.~\cite{{Kuramashi:2016lql},{Sasaki:2007gw}}.

%
%
\begin{table}[!t]
\caption{
Summary of simulation parameters in 2+1 flavor QCD simulations. See Ref.~\cite{Ishikawa:2015rho} for further details.
}\label{Tab:Summary_Sim}
\begin{tabular}{ccccccccc}
\hline
$\beta$& $L^3\times T$  & $C_{\rm SW}$ &  $\kappa_{ud}$  & $\kappa_{s}$ 
&  $a^{-1}$ [GeV] &  $(La)^3$ & $m_{\pi}$ [MeV] & $N_{\rm conf}$\cr
\hline
1.82 &$96^3 \times96$ & 1.11 & 0.126117  & 0.124812 
& 2.333(18)  & $\sim (8.1\; {\rm fm})^3$ & 145 & 200
\cr
\hline
\end{tabular}
\end{table}

\section{Electric and magnetic form factors}

\subsection{Numerical results}
\begin{figure}[h]
  \includegraphics[width=0.5\textwidth, clip]{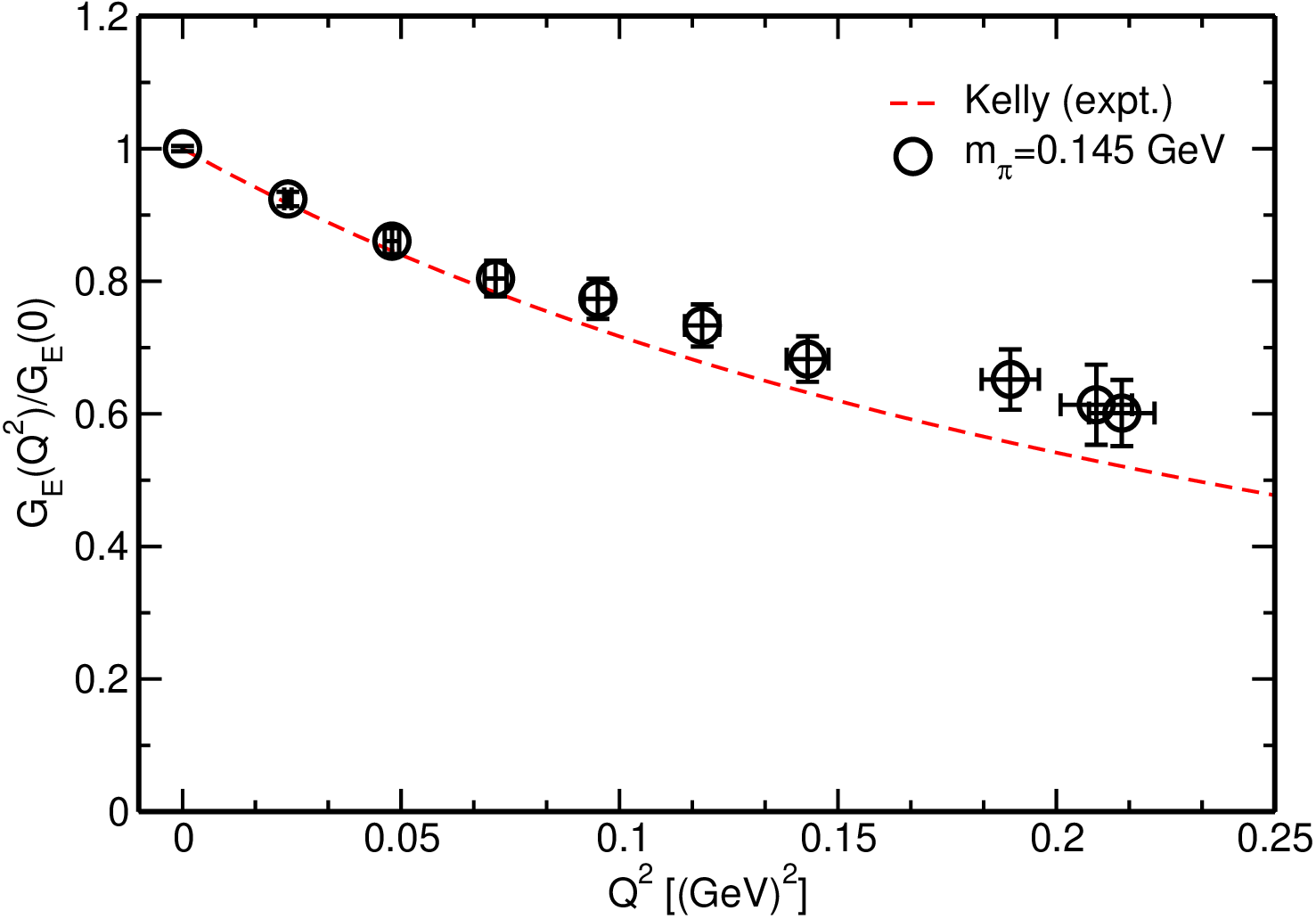}
  \includegraphics[width=0.49\textwidth, clip]{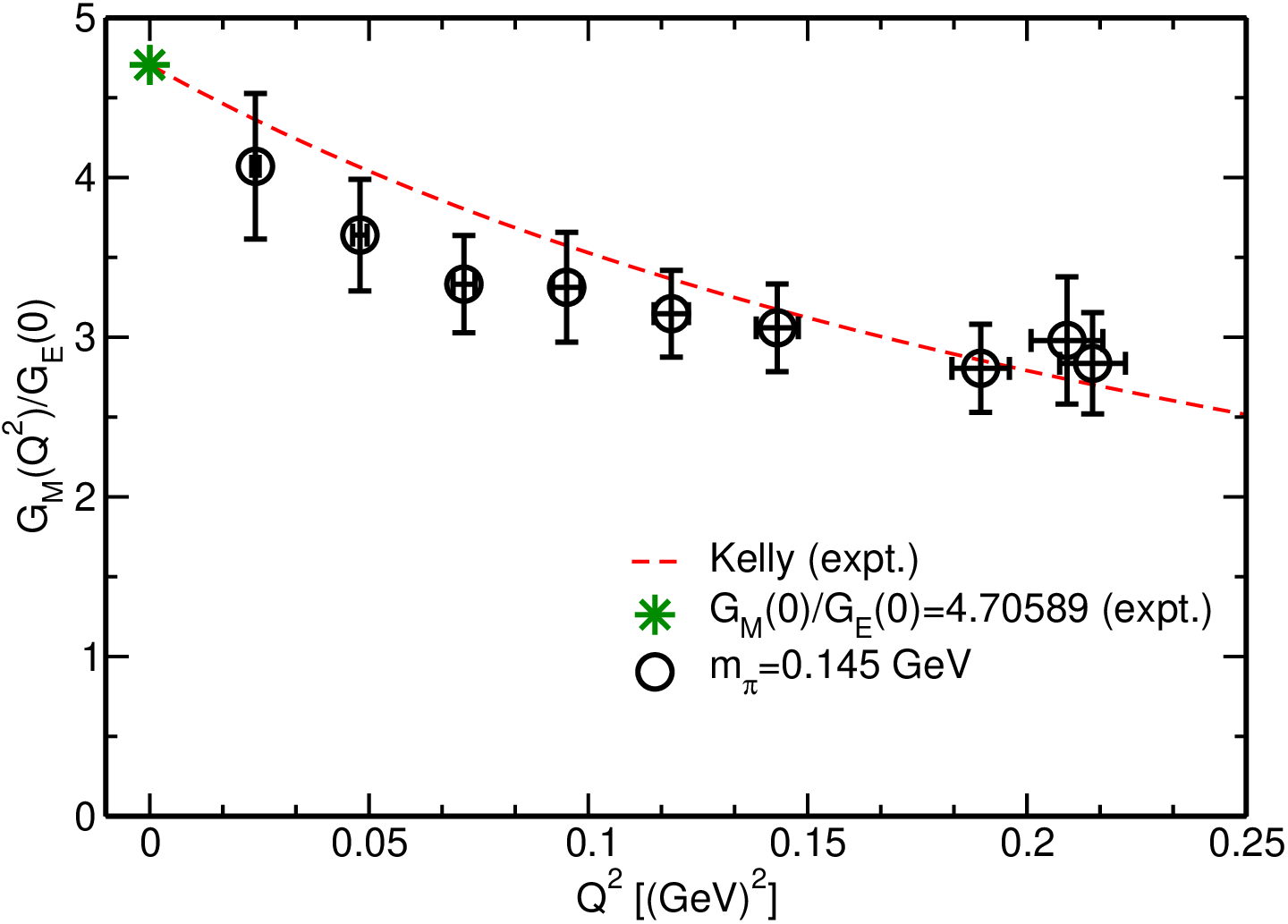}
  \caption{Momentum-squared dependence of the iso-vector electric form factor $G_E$ (left) and 
  magnetic form factor $G_M$ (right).\label{fig:gegm}}
\end{figure}

First of all, we show the results of iso-vector electric form factor $G_E$ (left panel)
and magnetic form factor $G_M$ (right panel) in Fig.~\ref{fig:gegm}. For the finite three momentum $\bm q$, 
we use the nine lowest nonzero momenta: ${\bm q}=2\pi/L\times {\bm n}$
with $|{\bm n}|^2\le 9$. The lowest nonzero momentum transfer in present work 
reaches the value of 0.024(1) $[({\rm GeV})^2]$, which is much smaller than
$4m_\pi^2$ even at $m_\pi=145$ MeV. In each panel, we also plot 
Kelly's fit~\cite{Kelly:2004hm} (red dashed curves) as their experimental data.
The simulated electric form factor is fairly consistent with the experiment,
especially at low $q^2$. The magnetic form factor agrees with the Kelly's fit albeit with
rather large errors. 

%
%

The electric and magnetic RMS radii can be evaluated from
the slope of the respective form factor at zero momentum transfer as described previously.
Recall that $G_M$ at the zero momentum transfer, whose value corresponds to
the iso-vector magnetic moment $\mu_V=G_M(0)$, cannot be directly measured for the kinematical 
reason~\cite{Sasaki:2007gw}. Therefore, $q^2$-extrapolation is necessary to evaluate the value
of $G_M(0)$.
In principle, low $q^2$ data points are crucial for determination of both the RMS radii and magnetic moment.
However, the accessible momentum transfer is limited on the lattice since the value of the squared three momentum transfers is discrete in units of $(2\pi/L)^2$. In this sense, a large spatial size $L$ is required for precise determination of the RMS radii ($r_E$ and $r_M$) and magnetic moment ($\mu_V$).

\subsection{Analysis with $z$-expansion method}
The determination of the slope (or $q^2$-extrapolation) is highly sensitive to how we fit the $q^2$-dependence of the
form factors. In the previous studies~\cite{Bhattacharya:2013ehc,Green:2014xba,Capitani:2015sba,Alexandrou:2013joa}, the dipole form $G(q^2) = a/(b+q^2)^2$ and the polynomial form 
$G(q^2)  = \sum_{k=0} a_kq^{2k}$ have been often adopted for $G_E$ and $G_M$. 
However, the fitting form ansatz may tend to constrain the interpolation (or extrapolation) 
and introduce a model dependence into the final result of the RMS radius (or the value of $G(0)$).
In order to reduce systematic errors associated with an interpolation of the form factor in momentum transfer, 
we use the model-independent $z$-expansion method~\cite{{Boyd:1995cf},{Hill:2010yb}}.

Let us recall the analytic structure of the form factors in the complex $q^2$-plane.
There is a non-analytic region for $G(q^2)$ due to threshold of two or more particles, \textit{e.g.} 
 $2\pi$ continuum. Hence the $q^2$-expansion, $G(q^2) = \sum_{k=0}a_kq^{2k}$, does not converge if
 $|q^2| > 4m_\pi^2$ where $q^2 = -4m_\pi^2$ is a branch point associated 
 with the pion pair creation for $G=G_E$ or $G_M$~\cite{Hill:2010yb}.
 The $z$-expansion (denoted as z-Exp) makes use
 of a conformal mapping from $q^2$ to a new variable $z$.
 Since this transformation makes the analytic domain mapped inside a unit-circle
 $|z| < 1$ in the $z$-plane, the form factors can be described by 
 a convergent Taylor series in terms of $z$:
 %
 %
 \begin{align}
  G(z) & = \sum_{k=0}^{k_{\rm max}} c_k z(q^2)^k & &  {\rm with} & z(q^2)=
  \frac{\sqrt{4m_{\pi}^2+q^2}-2m_\pi}{\sqrt{4m_{\pi}^2+q^2}+2m_\pi}.
 \end{align}
 where $k_{\rm max}$ truncates an infinite series expansion 
 in $z$ and $m_\pi$ corresponds to the simulated pion mass.
 Once the ratios of
 $|c_k/c_{k-1}|$ reach a convergence value less than unity at $k<k_{\rm max}$, 
 the z-Exp method makes {\it a model independent fit} and reduces
 systematic uncertainties to determine the RMS radius and the value of $G(0)$.

%
%
\begin{figure}[h]
  \includegraphics[width=0.5\textwidth,clip]{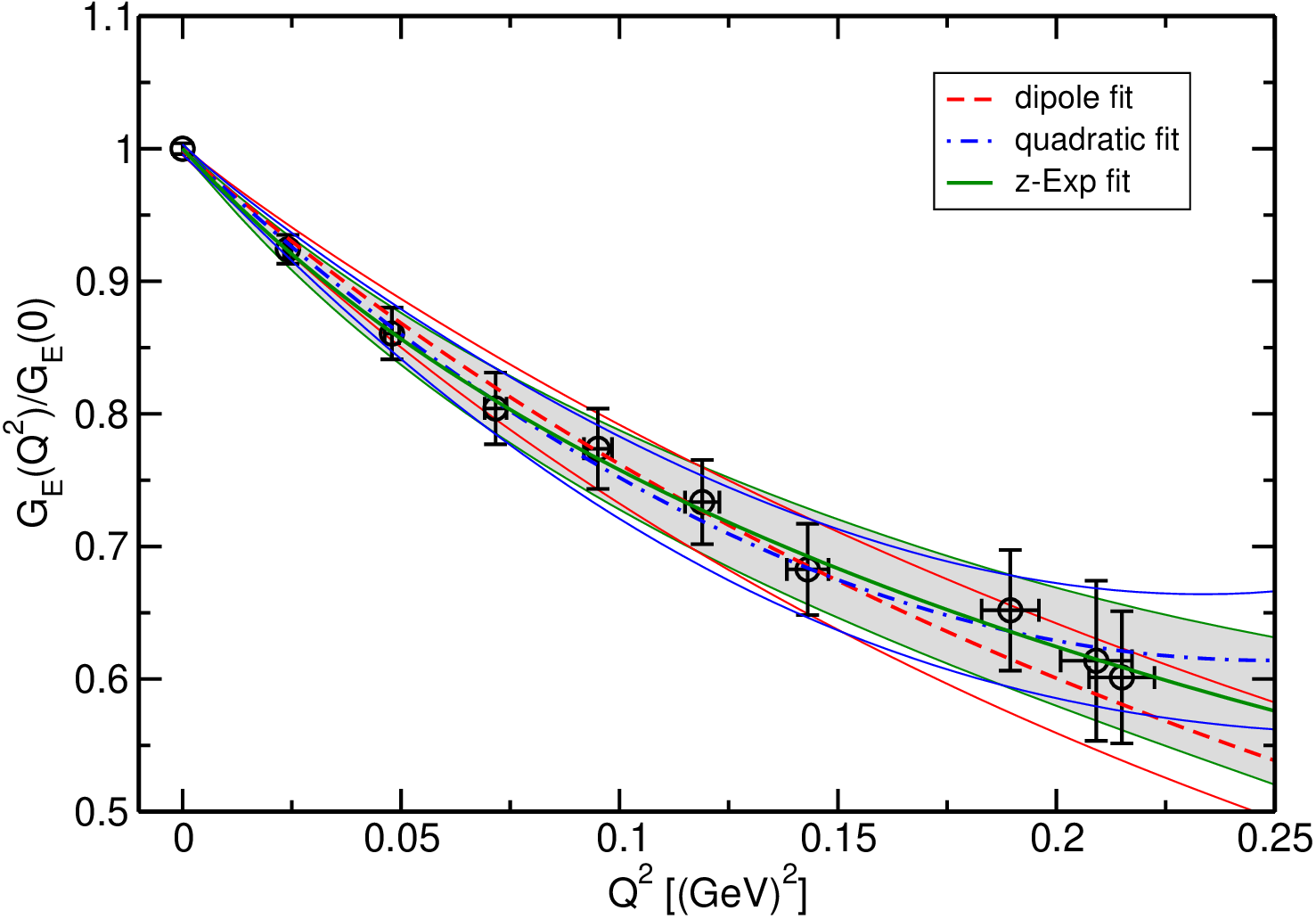}
     \includegraphics[width=0.49\textwidth,clip]{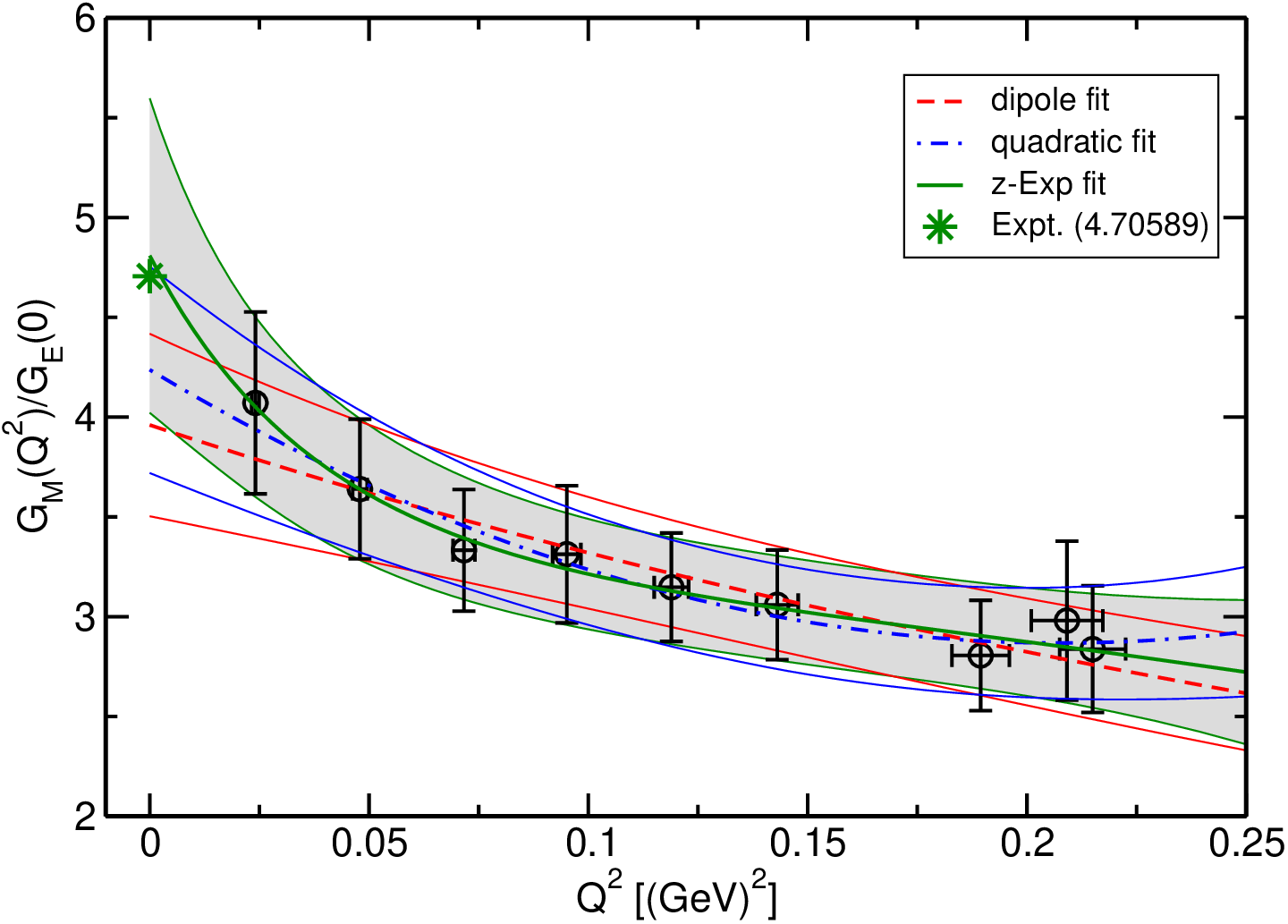}
     \caption{
     Results of $G_E$ (left panel) and $G_M$ (right panel)
     with three types of the fitting form ansatz: dipole (red),
     quadratic (blue) and z-Exp (green) fits.
     All fits are performed with all 10 (9) data points for $G_E$ ($G_M$).\label{fig:gegmfit}}
\end{figure}

In Fig.~\ref{fig:gegmfit}, we show the fit results of $G_E(q^2)$ (left panel) and $G_M(q^2)$ (right panel)
with three types of fits: dipole (red), quadratic (blue) and z-Exp (green) fits.
All the fits reasonably describe all 10 (9) data points for $G_E$ ($G_M$) with $\chi^2/{\rm dof}<1$.
However, if one takes a closer look at low $q^2$, the fit curve given by 
the $z$-expansion fit goes nicely through the data points in low $q^2$ region. 
This implies that the z-Exp fit is relatively insensitive on the higher $q^2$ data points
as was expected. Indeed, the z-Exp gives a rapid convergence series, which guarantees
that the fit result is observed to be insensitive on the choice of $k_{\rm max}$ if $k_{\rm max}\ge 3$. 
We then obtain $r_E = 0.917\pm0.099\;\mathrm{fm}$ from $G_E$ and $\mu_V = 4.81\pm0.79$
from $G_M$.

\begin{figure}[thb]
 \includegraphics[width=0.5\textwidth,clip]{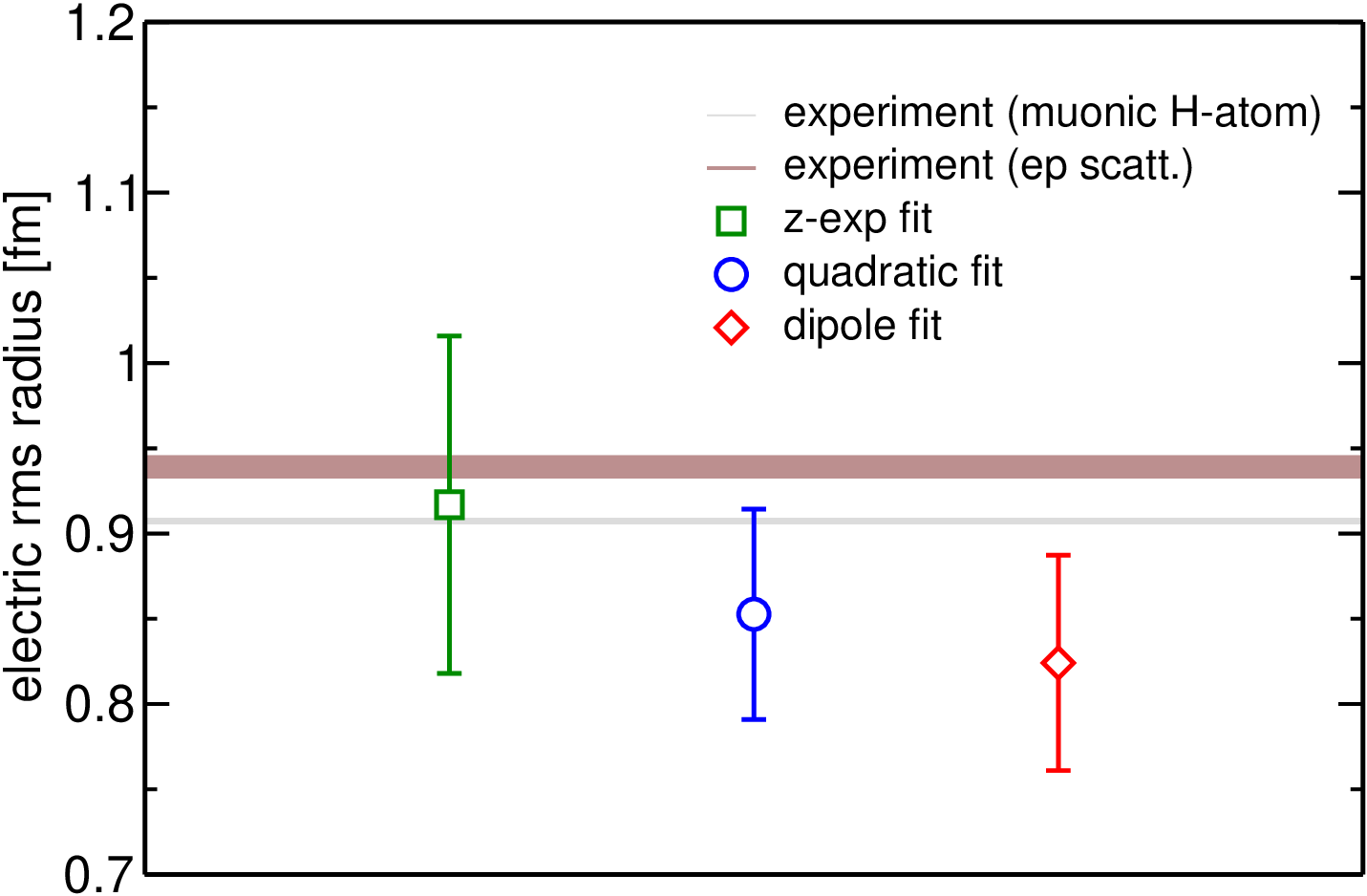}
 \includegraphics[width=0.49\textwidth,clip]{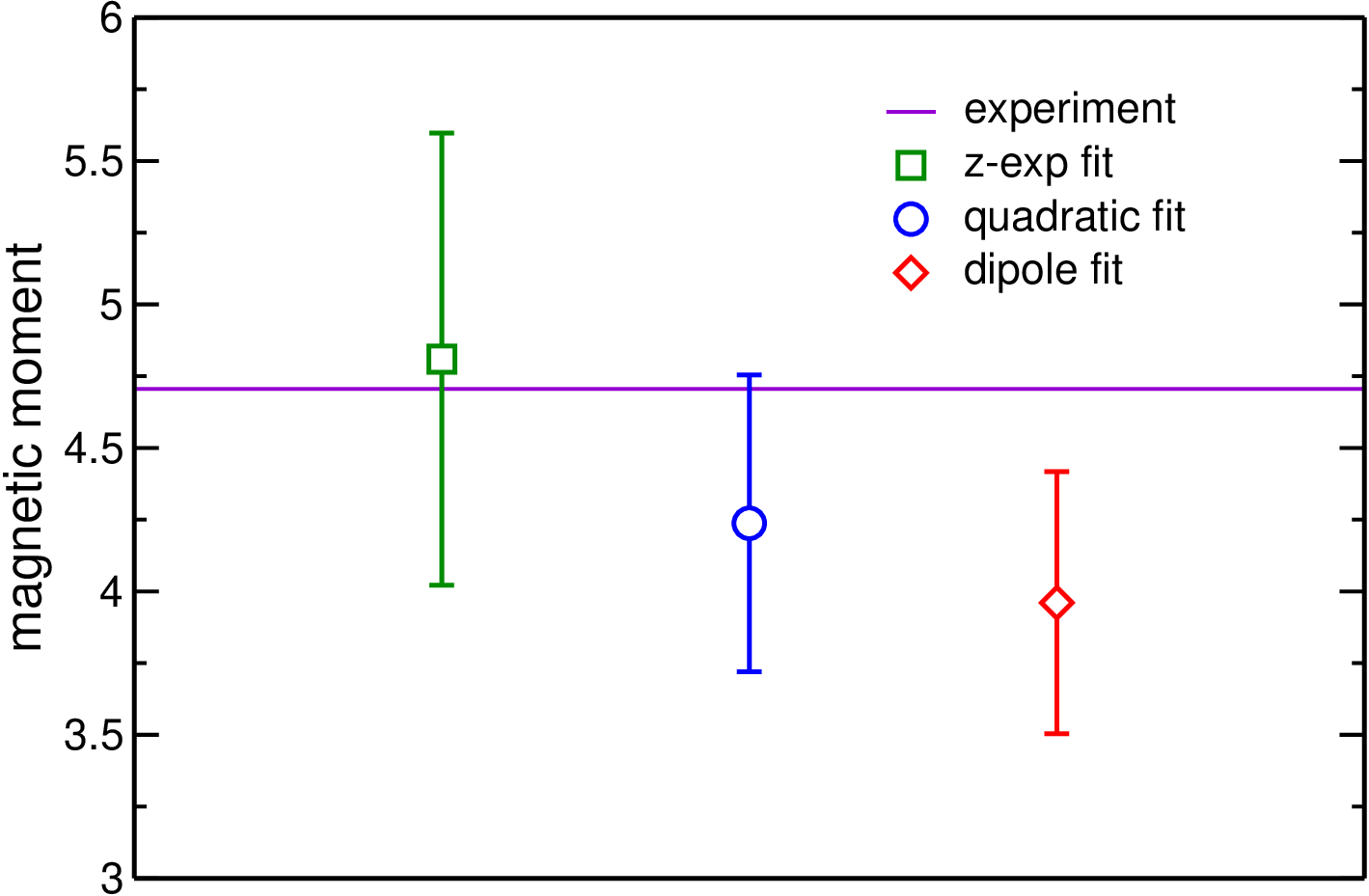}
 \caption{Comparison among results obtained with three types of the fitting form ansatz
 for the electric RMS radius $r_E$ (right panel) and magnetic moment $\mu_V$ (left panel).
 In the left panel, two horizontal bands represent experimental results from muonic $H$ atom (gray)
 and $ep$ scattering (brown).\label{fig:RMSaMM}}
\end{figure}

\begin{figure}[thb]
  \centering
  \sidecaption
 \includegraphics[width=0.5\textwidth,clip]{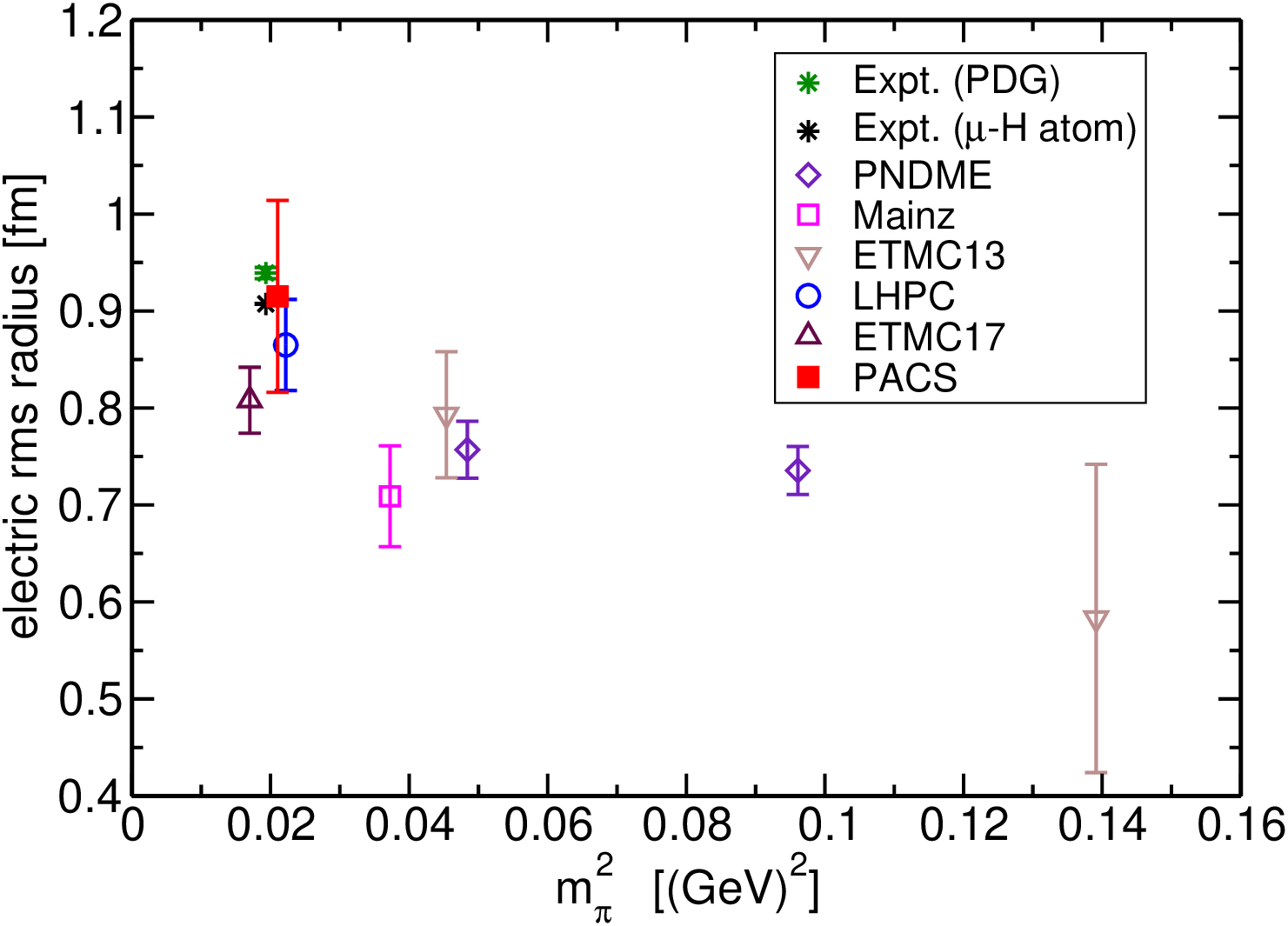}
 \includegraphics[width=0.49\textwidth,clip]{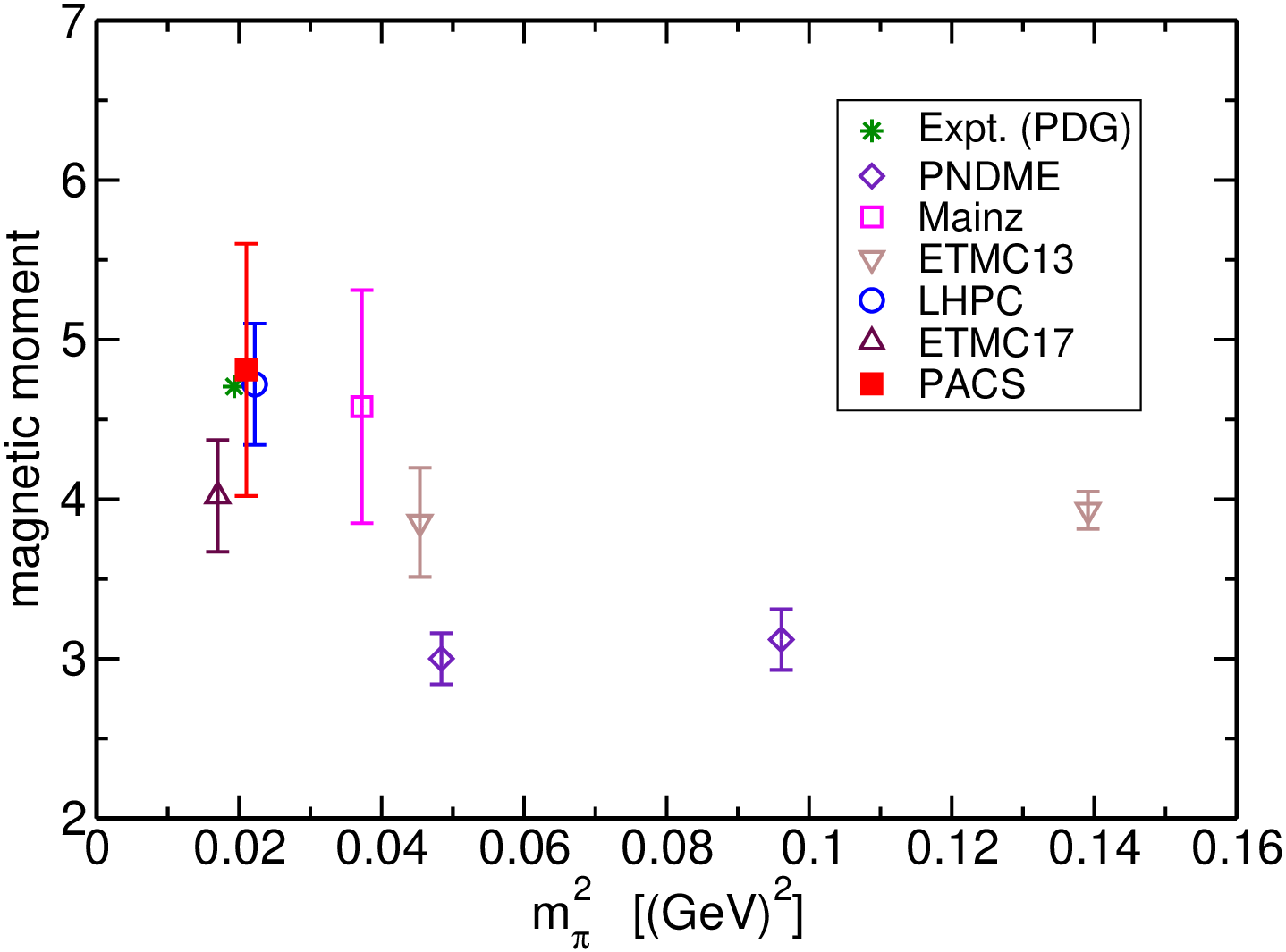}
 \caption{Our results for $r_E$ (left) and $\mu_V$ (right) at $m_\pi=145$ MeV
 (filled square) compared to recent lattice results~\cite{Bhattacharya:2013ehc,Green:2014xba,Capitani:2015sba,Alexandrou:2013joa,Alexandrou:2017ypw}.
 The asterisk symbols represent the experimental results~\cite{Patrignani:2016xqp}.
 \label{fig:RMS}}
\end{figure}

In Fig.~\ref{fig:RMSaMM}, we compare the results from the z-Exp fit with
those of dipole and quadratic fits. Although the z-Exp fit result has relatively 
larger error than the other results, its error may include both statistical and systematic uncertainties
without any model dependence. 
Moreover each result of $r_E$ and $\mu_V$ from the z-Exp fit 
is most consistent with respective experiment. We finally compare our results with 
recent lattice results from PNDME~\cite{Bhattacharya:2013ehc}, LHPC~\cite{Green:2014xba}, the Mainz group~\cite{Capitani:2015sba}, and ETMC~\cite{Alexandrou:2013joa,Alexandrou:2017ypw} as shown Fig.\ref{fig:RMS}.

%
%
%

%
%
%
\section{Axial vector form factors}
\subsection{Numerical results}
\begin{figure}[!h]
    \includegraphics[width=0.5\textwidth,clip]{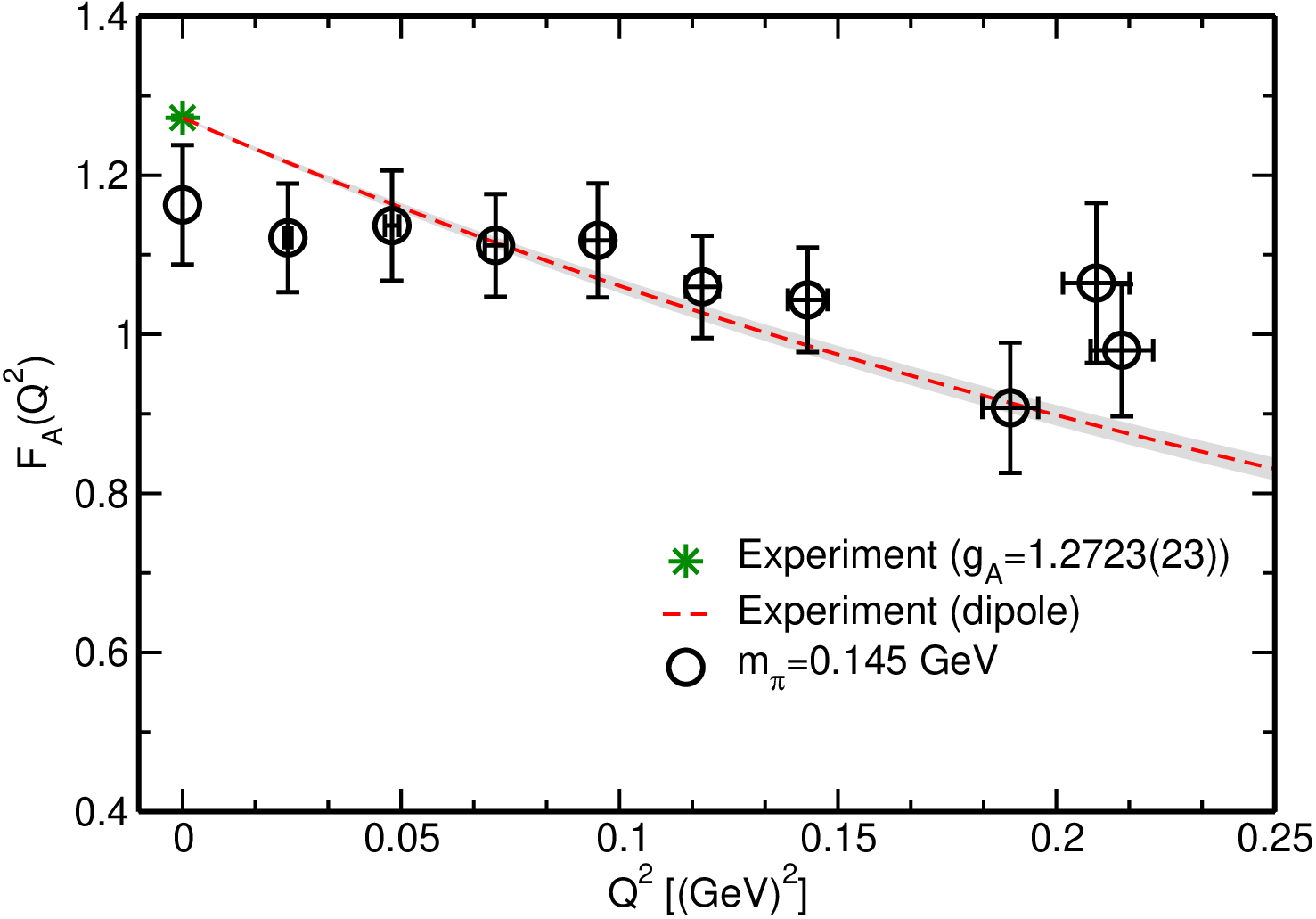}
    \includegraphics[width=0.5\textwidth,clip]{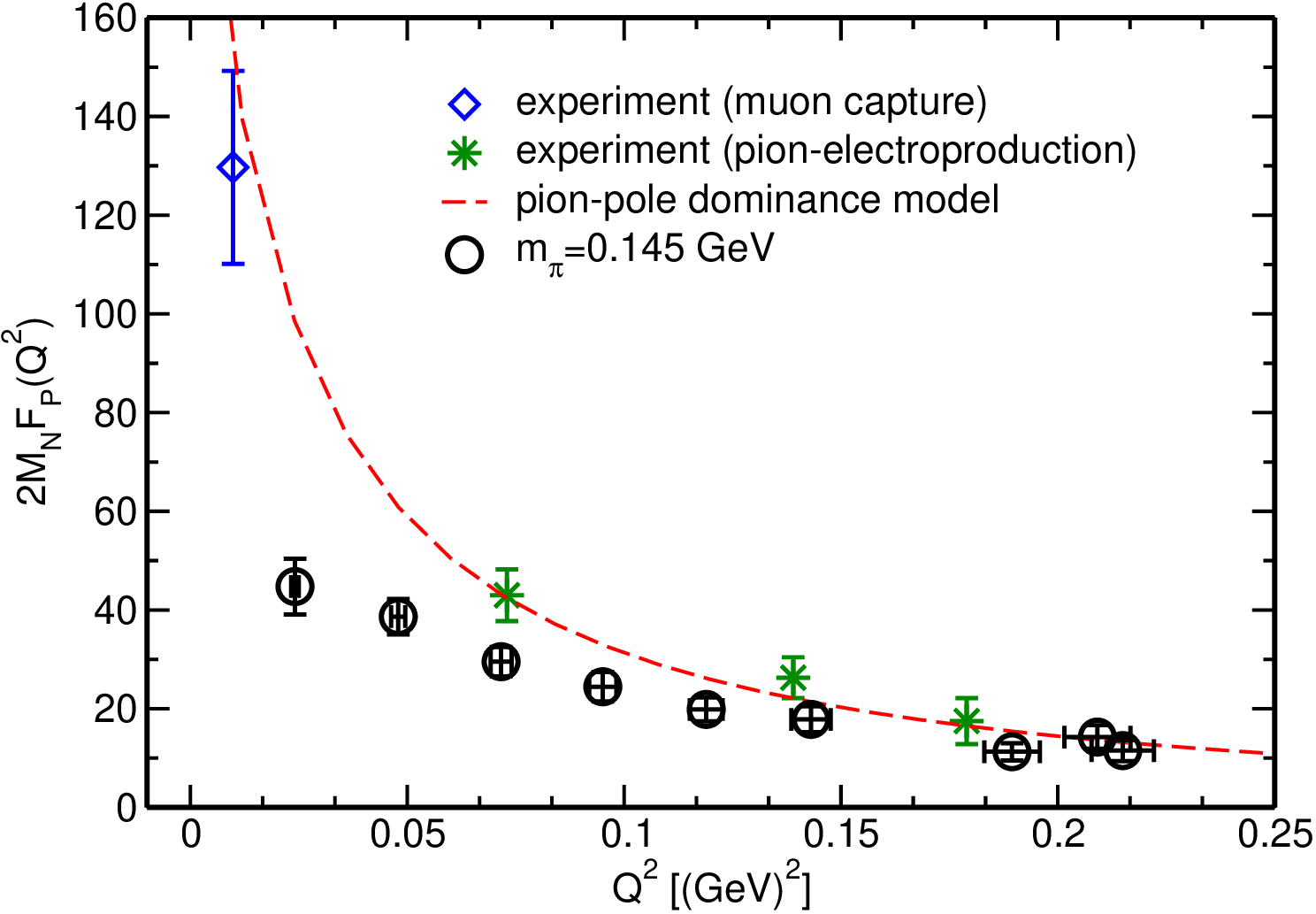}
    \caption{Momentum-squared dependence of the 
    axial-vector from factor $F_A$ (left) and induced-pseudoscalar form
    factor $F_P$ (right). Both form factors are renormalized with
    $Z_A=0.9650(68)$, which is given in the SF scheme~\cite{Ishikawa:2015fzw}. 
    In the left panel, the experimental curve is given by a dipole form with the RMS
    radius of 0.66(14) fm, while the red dashed curve in the right panel is given by
    the pion-pole dominance model. 
     \label{fig:fafp}}
\end{figure}

 We next show preliminary results for the nucleon form factors, $F_A$ and $F_P$, 
 in the axial-vector channel in Fig.~\ref{fig:fafp}.
 Although $F_A$ is barely consistent with the experimental curve, $F_P$ 
 is underestimated in comparison with both experiments and the pion-pole dominance (PPD) 
 model. Although two experiments of muon capture and pion-photo production
 follow a curve given by the PPD model, where the induced pseudo-scalar form factor 
 is given as $F_P^{\rm PPD}(q^2)=2M_N F_A(q^2)/(q^2+m_\pi^2)$, such functional
 form becomes justified by the generalized GT relation~(\ref{Eq:GTrelation}) only in the
 chiral limit ($\hat{m}=0$). However, Fig.~\ref{fig:fafp} may suggest that the measured
 $F_P(q^2)$ does not reproduce the expected pion-pole behavior.
 
%
%

\subsection{Test for the axial Ward-Takahashi identiy} 
\begin{figure}[thb]
     \sidecaption
     \includegraphics[width=0.5\textwidth,clip]{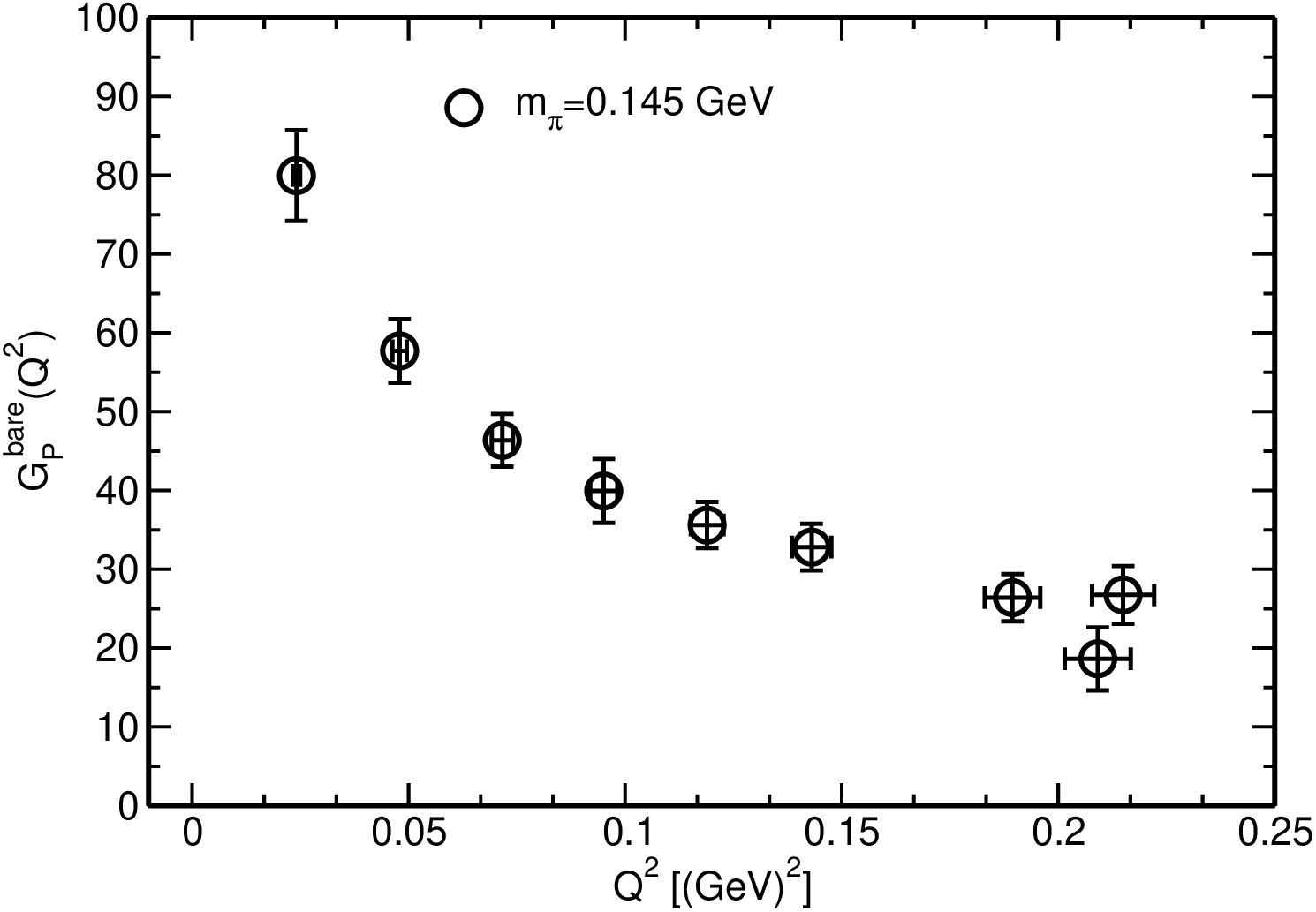}
     \caption{Momentum-squared dependence of the pseudo-scalar form factor, which is the bare value,
     {\it i.e.} not renormalized.\label{fig:gp}}
\end{figure}

\begin{figure}[thb]
     \sidecaption
     \includegraphics[width=0.5\textwidth,clip]{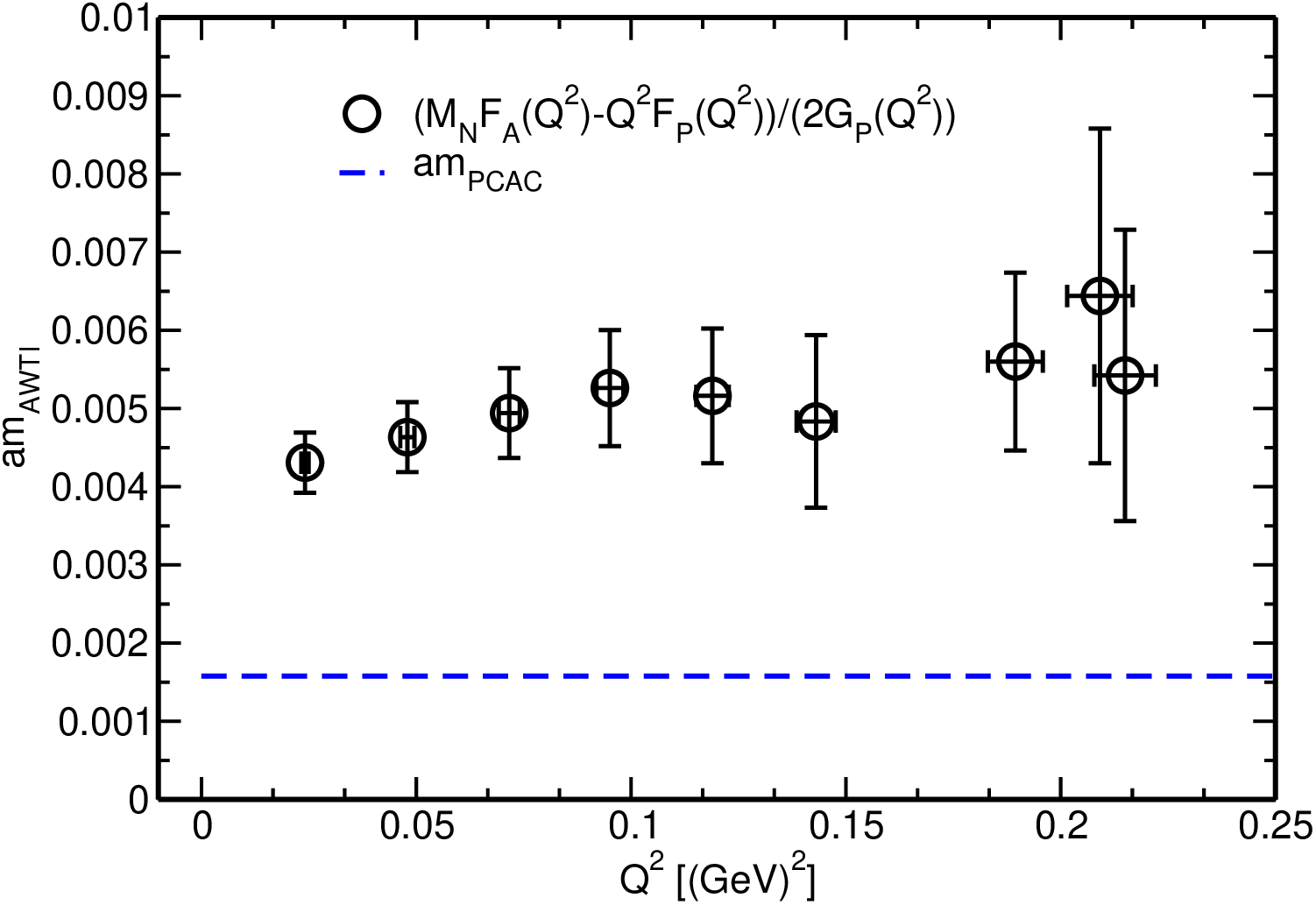}
    \caption{The ratio $m_{\rm AWTI}$ defined in Eq.~(\ref{Eq:MAWTI}) as a function of 
    momentum squared $q^2$.  A blue dashed line denotes the PCAC quark mass~\cite{Ishikawa:2015rho, Ishikawa:2015fzw}.
    \label{fig:awi}}
\end{figure}

 To test whether the correct chiral behavior is well satisfied in our simulations, 
 we also calculate the pseduo-scalar matrix element, which is described by
 the single form factor called as the pseduo-scalar form factor $G_P(q^2)$ defined in Eq.~(\ref{Eq:PSFF}). 
 In Fig.~\ref{fig:gp}, we plot the bare pseudo-scalar form factor $G_P(q^2)$, which is not 
 renormalized. The measured $q^2$ dependence of $G_P(q^2)$ resembles that of $F_P(q^2)$, where
 the relatively strong $q^2$ dependence appears in the lower $q^2$ region 
 as was expected from the pion-pole contribution.
 In order to verify the axial Ward-Takahashi identity in terms of the nucleon matrix elements,
 we define the following ratio inspired by the generalized GT relation~(\ref{Eq:GTrelation}):
%
%
\begin{align}
 m_{\rm AWTI} = \frac{2M_NF_A(q^2) - q^2 F_P(q^2)}{2G_P(q^2)}
 \label{Eq:MAWTI}
\end{align}
with the simulated nucleon mass $M_N$. If the ratio reveals a $q^2$ independent
plateau in the entire $q^2$ region, $m_{\rm AWTI}$
should be regarded as a bare quark mass. As shown in Fig.~\ref{fig:awi},
the ratio $m_{\rm AWTI}$ has no appreciable $q^2$ dependence. 
It indicates that three form factors well satisfy the generalized GT relation. 
We can read off $am_{\rm AWTI} \sim 0.005$, which is more than 3 times
larger than an alternative (bare) quark mass ($am_{\rm PCAC} =0.001577(10)$)
obtained from the pion two-point correlation functions with the partially conserved axial-vector 
current (PCAC) relation~\cite{Ishikawa:2015rho, Ishikawa:2015fzw}. 
Since the PCAC relation is also a consequence of the axial Ward-Takahashi identity, 
a relation $m_{\rm PCAC}\approx m_{\rm AWTI}$ was expected. However, it is not the case.
This issue may be connected with modification of the pion-pole structure in 
both $F_A(q^2)$ and $G_P(q^2)$. Such studies are now under way.

%
%
%

%
%
%

\section{Summary}
We have studied the nucleon form factors calculated in 2+1 flavor
QCD near the physical point ($m_\pi$ =145MeV) on a large spatial volume $(8.1\mathrm{fm})^3$.
We carefully examine both electric and magnetic form factor
shapes with a model-independent analysis based on the $z$-expansion method.
As a result, we obtain the iso-vector electric RMS radius $r_E$ and magnetic moment
$\mu_V$, which are consistent with experimental values.
We also analyzed axial and pseudoscalar nucleon matrix elements, which are described by
$F_A$, $F_P$ and $G_P$ form factors. We have found that the axial-vector form factor 
$F_A$ is barely consistent 
with experiments. In addition, $q^2$ dependences of three form factors,
$F_A, F_P$ and $G_P$ are well constrained by the generalized GT relation
as a consequence of the axial Ward-Takahashi identity, while the induced pseudo-scalar form 
factor $F_P$ is itself underestimated in comparison with experiments in low $q^2$ region.

\clearpage

\vskip3mm
\noindent
\textbf{Acknowledgments:} Numerical calculations for the present work have been carried out on the FX10 supercomputer system at Information Technology Center of the University of Tokyo, on the HA-PACS and COMA cluster systems under the ``Interdisciplinary Computational Science Program'' of Center for Computational Science at University of Tsukuba, 
on the Oakforest-PACS system of Joint Center for Advanced High Performance Computing,
on HOKUSAI GreatWave at Advanced Center for Computing and Communication of RIKEN, and on the computer facilities of the Research Institute for Information Technology of Kyushu University. 
This research used computational resources of the HPCI system provided by Information Technology Center of the University of Tokyo, Institute for Information Management and Communication of Kyoto University, the Information Technology Center of Nagoya University, and RIKEN Advanced Institute for Computational Science through the HPCI System Research Project (Project ID: hp120281, hp130023, hp140209, hp140155, hp150135, hp160125, hp170022). We thank the colleagues in the PACS Collaboration for helpful discussions and providing us the code used in this work. 
This work is supported in part by MEXT SPIRE Field 
5, and also by Grants-in-Aid 
for Scientific Research from the Ministry of Education, Culture, Sports, 
Science and Technology (No. 16H06002).

\end{document}